\documentclass{acm_proc_article-sp}%%%%where cta-author is the template name

\usepackage{graphicx}
\usepackage{epstopdf}
\usepackage{threeparttable}
\usepackage{multirow}
\usepackage[ruled,vlined]{algorithm2e}
\usepackage{amsopn}
\usepackage{diagbox}
\usepackage[table]{xcolor}
\usepackage{booktabs}

\newcommand{\tabincell}[2]{\begin{tabular}{@{}#1@{}}#2\end{tabular}} 

\begin{document}

\title{Perspectives on Stability and Mobility of Transit Passenger's Travel Behavior through Smart Card Data}

% \author{\au{Zhiyong Cui$^{1}$}%%% First author
% \au{Ying Long$^{2 *}$}%%% Second author
% %\au{J. Zhou$^2$}%%% Third author
% }

% \address{\add{1}{Department of Civil and Environmental Engineering, University of Washington, Seattle, WA 98195, USA}%%% Author address here
% %%% First group represent author affiliation number and second group represent name.
% \add{2}{School of Architecture, Tsinghua University, China}
% \add{*}{Corresponding Author}
% \email{zhiyongc@uw.edu, ylong@tsinghua.edu.cn}}

\numberofauthors{2}

\author{
% 1st. author
\alignauthor Zhiyong Cui\\
       \affaddr{Department of Civil and Environmental Engineering}\\
       \affaddr{University of Washington, Seattle, USA}\\
       \email{zhiyongc@uw.edu}
% 2nd. author
\alignauthor Ying Long\\
       \affaddr{School of Architecture}\\
       \affaddr{Tsinghua University, Beijing, China}\\
       \email{ylong@tsinghua.edu.cn}
}
\maketitle
\begin{abstract}
Existing studies have extensively used spatiotemporal data to discover the mobility patterns of various types of travelers. Smart Card Data (SCD) collected by the Automated Fare Collection (AFC) systems can reflect a general view of the mobility pattern of public transit riders. Mobility patterns of transit riders are temporally and spatially dynamic, and therefore difficult to measure. However, few existing studies measure both the mobility and stability of transit riders' travel patterns over a long period of time. In this paper, to analyze the long-term changes of transit riders' travel behavior, we define a metric for measuring the similarity between SCD. We also propose an improved density-based clustering algorithm, Simplified Smoothed Ordering Points To Identify the Clustering Structure (SS-OPTICS), to identify transit rider clusters. Comparing to the original OPTICS, SS-OPTICS needs fewer parameters and has better generalization ability. Further, the generated clusters are categorized according to their features of regularity and occasionality. Based on the generated clusters and categories, fine- and coarse-grained travel pattern transitions of transit riders over four years from 2010 to 2014 are measured. By combining socioeconomic data of Beijing in the year of 2010 and 2014, the interdependence between stability and mobility of transit riders' travel behavior is also discussed.
\end{abstract}

 % NOT

\section{Introduction}

The continuum of human spatial immobility-mobility at varying geographic and temporal scales poses fascinating topics and challenges for researchers and governments to make decisions on urban planning and transportation development. Stability and mobility are relevant, since mobility reflects the movement in short-term temporal or small spatial scales, while stability refer to long-term duration of stay or large scale locational consistency. Geographically, people move over scales ranging from a few meters to hundreds of kilometers in metropolitan areas; temporally, they move or stay over scales ranging from a few minutes to many years. Although people's movement seems to be disordered, we can still mining useful patterns for both individuals and a group of residents from various types of data. 

The temporal and spatial dynamic mobility pattern of residents has been concerned about for a long time by researchers in the fields of transportation engineering\cite{Ma20131}, computer science \cite{Zheng2014Urban}, urban planning \cite{long2016early}, or even socioeconomics \cite{Hanson2005Perspectives}. Along with the development of computer science and geographic information system (GIS), many new technologies and new types of data can be utilized to measure people's mobility pattern in large-scale regions, such as Call Detail Records of mobile phone \cite{Kang2013Exploring}, taxicabs' GPS information \cite{Peng2012Collective}, or even outdoor Wi-Fi signal data. When comes to the city-wide mobility analysis, smart card data (SCD) collected by Automated Fare Collection (AFC) systems may be a better choice, since AFC system are widely adopted by public transportation operators in most metropolitan areas \cite{Mohamed2014Understanding}.

AFC systems based on contactless smart cards are available for both city buses and metros to record the details of transaction information when passengers boarding or alighting. SCD contains fine-grained information not only about passengers' ID (smart cards' ID) and locations of boarding or alighting stations, but also transaction time and bus/metro lines. It is a great convenience to utilize SCD to depict passengers' daily, weekly or yearly travel profiles in large-scale regions covered by public transit systems. From an individual perspective, SCD can help record the passenger's travel records, reflect his/her social and economic characteristics, and even forecast his/her routine travel patterns. From a city perspective, SCD acting as a transportation probe can help estimate transportation conditions and provide new materials for intelligent transit systems and urban planning policy. A large number of transit behavioral studies based on SCD have been carried out and gained popularities. However, there are still rooms for improvement in existing studies on SCD-based travel pattern analysis, which can be summarized into three aspects: 1) The SCD as a data source for mobility pattern research has inherent shortcomings; 2) The existing analysis methods also have drawbacks in terms of model efficiency and complicatedness; 3) Few studies focus on the long-term evolutionary travel behavior analysis. These shortcomings are described in detail in the following paragraphs and solved in this study.

Firstly, although SCD has multiple advantages, the shortcomings of SCD are obvious. The anonymous attribute of SCD determines the absence of basic personal information, like age and gender, without which it is hard to measure passengers' socioeconomic characteristics. For analyses spanning a long period of time, it may encounter the problems of changing smart cards and possessing multi-cards of passengers. In addition, some urban transit systems only record when and where passengers board transit buses and neglect the alighting information, leading to the extreme difficulty of inferring the destinations of passengers. Further, due to the increasing volume of rural-urban migration and the transient mobility of the internal migration in megalopolises, the measurement of the stability of the passenger's travel behavior will be highly influenced. Thus, for the sake of accurately measuring the stability and mobility of the transit passenger's travel behavior, effective classification or clustering methods should be employed.

However, existing SCD-based travel behavior analysis methods also have drawbacks. Since SCD normally does not contain the information for labeling the data, such as smart card owner information, classification methods can hardly be applied without labeling data. Hence, most existing studies employed clustering methods. With the help of clustering methods, transit travelers can be clustered into multiple groups, within which the grouped travelers share similar travel patterns. Most clustering methods, such as K-means, need to specify the number of clusters, i.e. the value of K, in advance. However, for the SCD clustering task, due to irregularity and variability of the passengers' travel behaviors, it is hard to decide how many clusters should be contained in the smart card data set. Some density-based clustering methods with no need to specify the number of  clusters, such as the Density-Based Scanning Algorithm with Noise (DBSCAN) and Ordering Points To Identify the Clustering Structure (OPTICS), have been applied for SCD clustering. But these methods normally need extra efforts to pre-set some meta parameters for these models, such as the distance threshold between clusters. Thus, to make the clustering process more efficient and convenience, we propose a simplified smooth OPTICS clustering methods to group the SCD in this study.

Much SCD-based work \cite{ma2017understanding,zhang2016spatiotemporal,mohamed2017clustering} focusing on mining transit passenger travel patterns attempted to distinguish the commuting trips, including non-transfer and with-transfer trips, and non-commuting trips. These various types of grouped trips are important to realize many short-term applications, such as travel time prediction and transportation scheduling. Even though existing studies have been conducted to characterize long-term urban dynamics using SCD \cite{long2015profiling}, how do travel patterns change over the years has not been well-studied. The long-term travel pattern changes are capable of reflecting the stability and mobility patterns of transit passengers and revealing the transportation development and the underlying urban evolution. 

In this study, we utilize the temporal information of SCD to mine the relationship between passenger's mobility and stability in different time and frequency scales. To overcome the drawbacks of the SCD, we define a metric for measuring the distance between different SCDs to better describe their similarity. Further, we propose an advanced density-based clustering method to group transit trips into different clusters. Based on the clustering results, we utilize SCD collected from different years to characterize the long-term stability and mobility of transit passengers' travel patterns. To better understand the passenger behavior in public transportation, we introduce other socioeconomic data into our analysis. Our contributions can be described as follows:
\begin{itemize}
\item We define an SCD similarity metric for measuring the difference between passengers' travel behaviors. To better describe the similarity, both the temporal difference and the frequency difference between SCD records are considered.
\item We propose a simplified smoothed OPTICS (SS-OPTICS) clustering method to cluster SCDs. Comparing to the classical OPTICS methods, the SS-OPTICS needs less parameters. We discover groups of passengers behaving similarly with respect to their boarding time.
\item We cluster the SCD based on different grouping granularities to analyze the long-term mobility and stability of transit passengers. To measure the evolutionary changes in the clustering results, socioeconomic data in different years is also incorporated in our analysis.
\end{itemize}

The organization of this paper is as follows: Related work is briefly discussed in section 2. In section 3, we proposed the SS-OPTICS algorithm and describe our methodology. Our analysis of mobility and stability is present in section 4. Section 5 concludes the paper with a summary and a short discussion of future research. 

\begin{figure}[t]
\centering
\includegraphics[width=\linewidth]{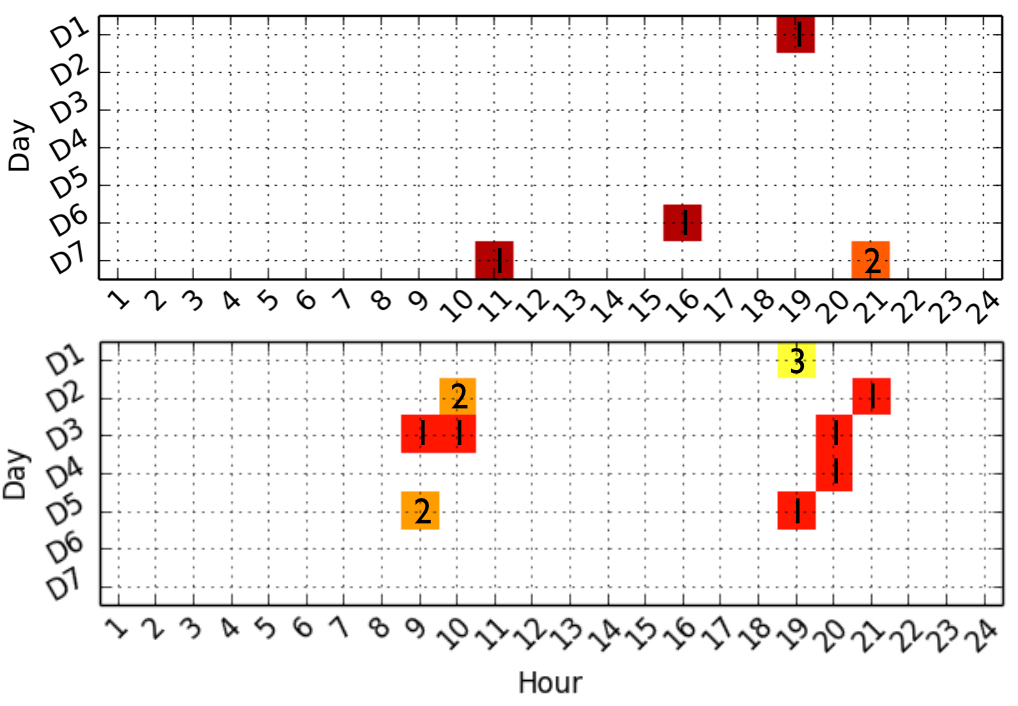}
\caption{Weekly profiles of two passengers' transaction time. The transaction time (colored squares) reflects their different travel patterns. The values in colored squares represent the number of transactions in that specific hour. D1-D5: weekdays, D6: Saturday, D7: Sunday.} 
\label{figure:transaction}
\end{figure}

\section{Related Work}

The concepts of stability and mobility of travel patterns are opposite but relevant. With the regard of the travel pattern analysis, the mobility tends to reflect the variations of passengers' travel patterns, while the stability characterizes the steadiness of patterns over a period of time. Hanson \cite{Hanson2005Perspectives} was among the first researchers to focus on stability analysis and stated that analyzing individuals' stability also requires analyzing their mobility. Through an empirical example centered on the relationship between entrepreneurship and place, he explicitly proposed that considering locational stability requires examining stability and mobility in tandem, since spatiotemporal dynamics involved. Based on this idea, James et al. \cite{Bagrow2012Mesoscopic} concentrated on detailed substructures and spatiotemporal flows of mobility to show that individual mobility is dominated by small groups of frequently visited, dynamically close locations, forming primary "habitats" capturing typical daily activity. While many other works \cite{Ma20131}, \cite{Noulas2012A}, \cite{Uppoor2014Generation}, \cite{Veloso2011Sensing} chose a perspective on large-scale mobility about urban human beings, vehicles or taxis.

To measure residents' stability and mobility in the urban area, SCD in public transit is one of the most widely used data. According to Long et al.\cite{Long2015Urban}, SCD related research topics can be classified as: 1) data processing and data complementation, such as back-calculation of origin and destination and recognition of trip purpose; 2) supporting and management of public transit systems; 3) place-based urban spatial structure and 4) person-based analysis on the social network and special group of people. Pelletier et al. \cite{Pelletier2011557} also gave a literature review of SCD usage in public transit and presented three levels of management of SCD: strategic (long-term planning), tactical (service adjustments and network development), and operational (ridership statistics and performance indicators). Zheng et al. \cite{Zheng2014Urban} presented several typical applications based on SCD, like building more accurate route planners. Further, Long et al.\cite{long2016early} sought to understand extreme public transit riders in Beijing using both traditional household surveys and SCD. In their work, public transit riders were classified into four groups of different types of extreme transit behaviors to identify the spatiotemporal patterns of these four extreme transit behaviors. Further, Neal et al. \cite{Lathia2013Individuals} discussed personalizing transport information services based on SCD. Among their contributions, the authors used clustering methods to prove that the usage of public transportation can vary considerably between individuals. Each passenger's trips were aggregated into a weekday profile describing his temporal habits and hierarchical agglomerative clustering is introduced to discover groups of passengers characterizing different travel habits. Contrary to this approach, our weekly profile, presented in Section 3, consisting of hour-grained grids can show more details. 

As we investigated, many clustering methods were adopted to process and analyze SCD. To clustering the temporal information, Mahrsi et al. \cite{Mohamed2014Understanding} constructed temporal passenger profiles based on boarding information and applied a generative model-based clustering approach to discover clusters of passengers. Based on the boarding information, passengers were assigned with "residential" areas, established through the clustering of socioeconomic data, to inspect how socioeconomic characteristics are distributed over the passenger temporal clusters. To analyze year-to-year changes in public transport passenger behavior, Briand et al. \cite{briand2017analyzing} propose a two-level generative model that applies the Gaussian mixture model to regroup passengers based on the passengers' temporal habits in their public transit usage. A density-based clustering method, DBSCAN \cite{DBSCAN1996Density}, which is very similar to OPTICS \cite{Ankerst1999OPTICS} is used by \cite{Ma20131}. The authors identified trip chains to detect transit riders' historical travel patterns and apply K-Means++ clustering algorithm and the rough-set theory to cluster and classify travel pattern regularities. To detect and update the daily changes in travel patterns, a Weighted Stop based DBSCAN is also proposed to reduce computation complexity \cite{kieu2015modified}. To achieve better clustering performance, a two-step clustering method \cite{mohamed2017clustering} was proposed to cluster transit stations and passengers, respectively. Comparing to approaches presented in these works, we improve the OPTICS algorithm to cut down input parameters and control cluster size. Further, other than focusing on people's mobility pattern, we utilize SCD to measure the interdependence between stability and mobility in the time dimension.

\begin{figure*}[t]
\centering
\includegraphics[width=0.8\linewidth]{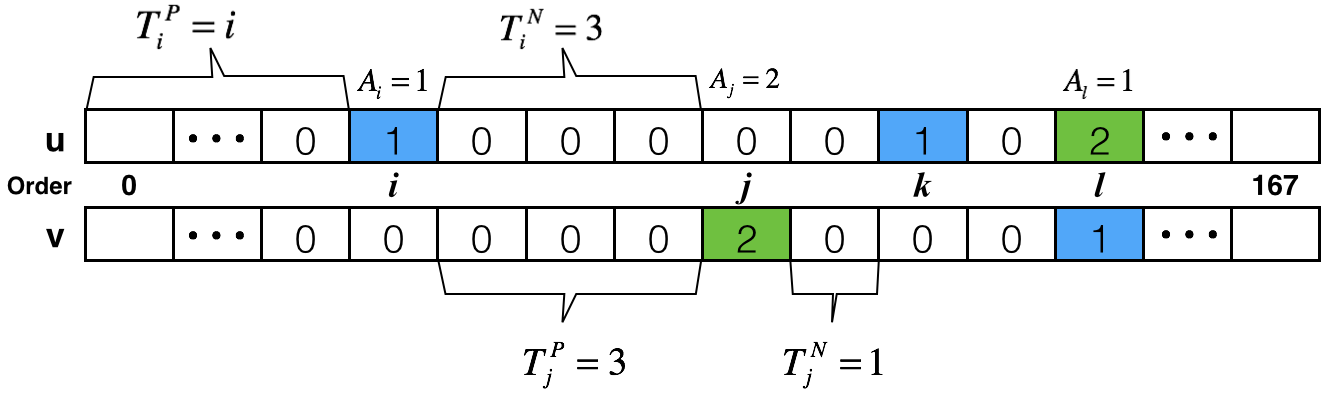}
\caption{Example of the distance between two vectors, \textbf{\emph{u}} and \textbf{\emph{v}}}
\label{figure:distance}
\end{figure*}

\begin{table}[b]
\centering
\caption{Definitions of extreme travelers \cite{long2016early}}
\begin{center}
\begin{tabular}{|c|c|}
\hline
Type & Definition  \\ \hline
\tabincell{c}{Early Birds \\(EBs)} & \tabincell{c}{First trip < 6AM, more than two days \\in five weekdays (60\% of weekdays)}  \\ \hline
\tabincell{c}{Night Owls \\(NOs)} & \tabincell{c}{Last trip > 10PM, more than two days\\ in five weekdays (60\% weekdays)} \\ \hline
\tabincell{c}{Tireless \\Itinerants (TIs)} &\tabincell{c}{ >= one and a half hours commuting, \\more than two days in a week} \\ \hline
\tabincell{c}{Recurring \\Itinerants (RIs)} & \tabincell{c}{>= 30 trips in weekdays of a week \\(>= 6 trips per day)} \\ \hline
\end{tabular}
\end{center}
\label{table:extreme}
\end{table}

\section{Profiling Passengers Based on SCD}

The transit passengers profiling process consists four stages: 1) pre-processing the SCD based on existing studies; 2) defining the distance (similarity) between different SCD records; 2) clustering samples of SCD with a proposed Simplified Smoothed OPTICS algorithm; and 3) classifying the whole SCD records with a K-means-like algorithm according to results of the clustering stage.

\subsection{Dataset Description}
The smart card data collected and issued by Beijing Transit Incorporated contains transit riders' records for both the bus and metro systems. There were two types of AFC systems on Beijing buses: flat fares and distance-based fares, before the beginning of 2015, since when all bus lines became distance-based fare systems. It is a design flaw for the bus smart card system that flat fares system records the transaction (paying) time when checking-in, whereas distance-based fares system records the transaction time when checking-out. For the Beijing metro system, although passengers pay the fare when alighting, the system records the time of both checking-in and checking-out. In this paper, to offset the design flaw, we consider the transaction time as the time for one ride.

We select SCD with shared card IDs from two datasets in 2010 and 2014. Both the selected datasets of 2010 and 2014 last for one week and contain the same smart card IDs with the amount of 1.9 million, representing 1.9 million passengers lived in Beijing at least from 2010 to 2014. We assume each smart card represents an anonymous passenger, without considering the situation of passengers' changing card, which is not common in Beijing. Each record of the SCD consists of 1) smart card ID, 2) boarding or alighting time, and 3) station ID of boarding or alighting line. As the time spans of SCD in 2010 and 2014 both cover one week, we estimate each passenger's trip activities using a "weekly profile", a vector contains 168 (7$\times$24) variables describing the distribution of the trip activities. Each variable in the vector represents the number of smart card's transaction time over each hour in each day of the week. Fig. \ref{figure:transaction} illustrates weekly profiles of passengers' transaction time.

\subsection{Data Pre-processing}
Before analyzing the transit behaviors of the passengers in Beijing, we separate the whole passengers into two groups: extreme travelers and non-extreme travelers, according to an existing study \cite{long2016early}. Four types of extreme travelers are defined based on their behaviors in weekdays, by setting several validated thresholds and combining empirical knowledge of Beijing as depicted in Table \ref{table:extreme}. For example, since most people's working hours start at 8:30 or 9:00 am in Beijing, public transit boarding time before 6:00 am would be considered as an unusually early situation \cite{long2016early}. As we evaluated, those types of extreme travelers only account for a small proportion (less than 5\%) of the whole passengers. Since the extreme travelers have clear definitions and can be easily filtered out from the raw data, the data of extreme travelers is eliminated from the dataset for clustering and analyzed separately in the experiments in our study.

\subsection{Definition of Distance between Smart Card Records}
After counting the number of transaction time of each smart card record, the feature of each record forms a travel record vector with 168 (24$\times$7) elements,  $\textbf{\emph{v}}=[v_0, \cdots , v_{167}]$, representing the number of smart card transactions in each hour of each day in the study period (one week). Each element of $\textbf{\emph{v}}$ is a non-negative integer that $\textbf{\emph{v}}_i \in \mathbb{Z}$. There are some classic distance-measurement methods to measure the similarity of different records, such as Euclidean distance, Manhattan distance, cosine distance, and cross correlation distance. However, since each transaction vector not only records the number of transactions but also contains temporal attributes, those classical distance formulas are not capable of comprehensively measuring the difference, i.e. the distance, between smart card transactions. 

The Euclidean distance can only measure the straight-line distance between two points, represented by two vectors with 168 dimensions in this study. Since the order of dimensions in the Euclidean space does not influence the distance, the Euclidean distance inherently misses out the temporal information between the SCD records. Similar to the Euclidean distance, the Manhattan distance measures the grid-based distance between two points and fails to consider the temporal information. The cosine distance, i.e. the cosine similarity, measures the similarity between two non-zero vectors of an inner product space and output the cosine of the angle between those two vectors. The cosine distance, commonly used in high-dimensional positive spaces, has already been used in SCD analysis \cite{zheng2014efficient}. However, since the travel record vectors are mostly sparse, the calculation of cosine distance conducting inner products on the two compared vectors will significantly cancel out the difference between the elements in the two vectors, if one of the elements in a vector is zero. For example, for the two travel record vectors \textbf{\emph{u}} and \textbf{\emph{v}}, if $v_i=0$ and $u_i\neq 0$, the cosine distance of \textbf{\emph{u}} and \textbf{\emph{v}} at the $i$-th position $\propto v_i*u_i = 0$, which still neglect the travel frequency difference at the $i$-th time point. The cross correlation distance (CCD) has also been used to measure similarity/distance between two sequences/vectors by shifting one sequence to find a maximum correlation with another sequence \cite{he2018classification}. However, due to the shifting mechanism, the calculation process of CCD almost gets rid of the relative position information of elements in two sequences. That means CCD cannot capture the difference of occurrence time of smart card transactions, and thus, CCD is not utilized in this study.

To solve this problem, we propose a new distance metric between two SCD transaction record vectors, defined as \textbf{Transaction Distance} (TD), to measure the difference between two transit passengers' travel patterns. Since most passengers normally do not take transit very frequent, most of the elements in a travel record vector are zeros. In this case, the non-zero values in the travel record vector greatly impact the difference between different vectors. We define the TD between the two vectors, \textbf{\emph{u}} and \textbf{\emph{v}}, by considering both the time difference and the frequency difference. Thus, the proposed TD consists of two parts, the riding time interval ($T_{i}$) and the absolute riding frequency difference ($A_{i}$), for each element $i$.
 
The absolute riding frequency difference is defined as $A_{i}=|u_{i}-v_{i}|$. As for the riding time interval $T_{i}$, if one of $u_{i}$ and $v_{i}$ equals to 0, $T_{i}$ equals the smaller value of $T_{i}^{P}$ and $T_{i}^{N}$, namely $T_{i} = \min \{T_{i}^{P}, T_{i}^{N}\}$. Here, taking $u_{i} \neq 0$ for example, $T_{i}^{P}$ represents the time interval between the current element $u_{i}$ and the nearest \textbf{previous} non-zero element in vector \textbf{\emph{v}}. And correspondingly, $T_{i}^{N}$ represents the time interval between the current element and the nearest \textbf{next} non-zero element in vector \textbf{\emph{v}}. If $u_{i}$ and $v_{i}$ both equal to or do not equal to 0, $T_{i}=0$. Fig. \ref{figure:distance} shows an example of how to compute the transaction distance that $T_{j} = \min\{T_{j}^{P}, T_{j}^{N}\}=1$ and $T_{l} = 0$. If a non-zero component in one vector cannot find a previous or next non-zero component in the other vector, like the situation of $u_{i}$ in Fig. \ref{figure:distance}, its $T_{i}^{P}$ equals $\min\{i,167-i\}$.
 
Then, the Transaction Distance between vectors \textbf{\emph{u}} and \textbf{\emph{v}} can be represented as:
\begin{equation}
TD = \sum_{i=0}^{167} \min \{ T_{i}^{P}, T_{i}^{N}\} + k*|u_{i}-v_{i}|, \quad s.t. \quad u_{i} \neq v_{i}
\end{equation}
Here, \emph{k} is a parameter to balance the weights of \emph{T} and \emph{A}. It is suitable for setting the value of \emph{k} ranging from 0 to 3, as we tested in the clustering section. 

\begin{figure}[t]
\centering
\includegraphics[width=\linewidth]{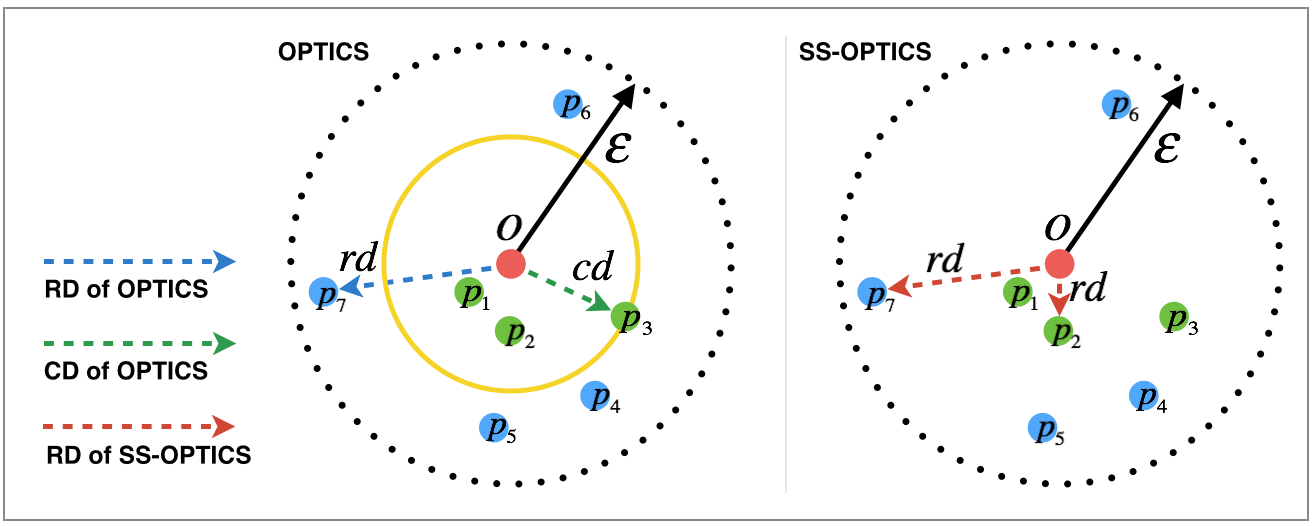}
\caption{The cd of OPTICS and the rd of both OPTICS and SS-OPTICS. \emph{MinPts} = 4}
\label{figure:OPTICS}
\end{figure}

\subsection{Clustering Samples of SCD Records}

After defining the distance between vectors of smart card records, we cluster the vectors to identify the travel patterns of public transit riders in Beijing. In order to accurately cluster the travelers, a suitable algorithm is needed. Although K-Means algorithms or other centroid based clustering models are very efficient, they need to nominate the number of clusters (K) before running of the algorithm. Even though iteratively setting K as different numbers and evaluating the performance may help to identify the proper value of K, dealing with high number of observations during this iterative process is still a big problem for K-Means. As the travel record vector has a high dimension (168) in this study, grid-based clustering methods are also not suitable for this problem. A new density-based clustering algorithm, clustering by fast search and find of density peaks \cite{Rodriguez2014Clustering}, is also tested. However, it can only identify 4 or 5 obvious clusters. Thus, to solve the aforementioned problems, we propose an improved density-based clustering algorithm based on OPTICS \cite{Ankerst1999OPTICS}, which is suitable for clustering the data based on the aforementioned transaction distances. We name it as the Simplified Smoothed OPTICS (SS-OPTICS). 

\begin{figure}[t]
\centering
\includegraphics[width=\linewidth]{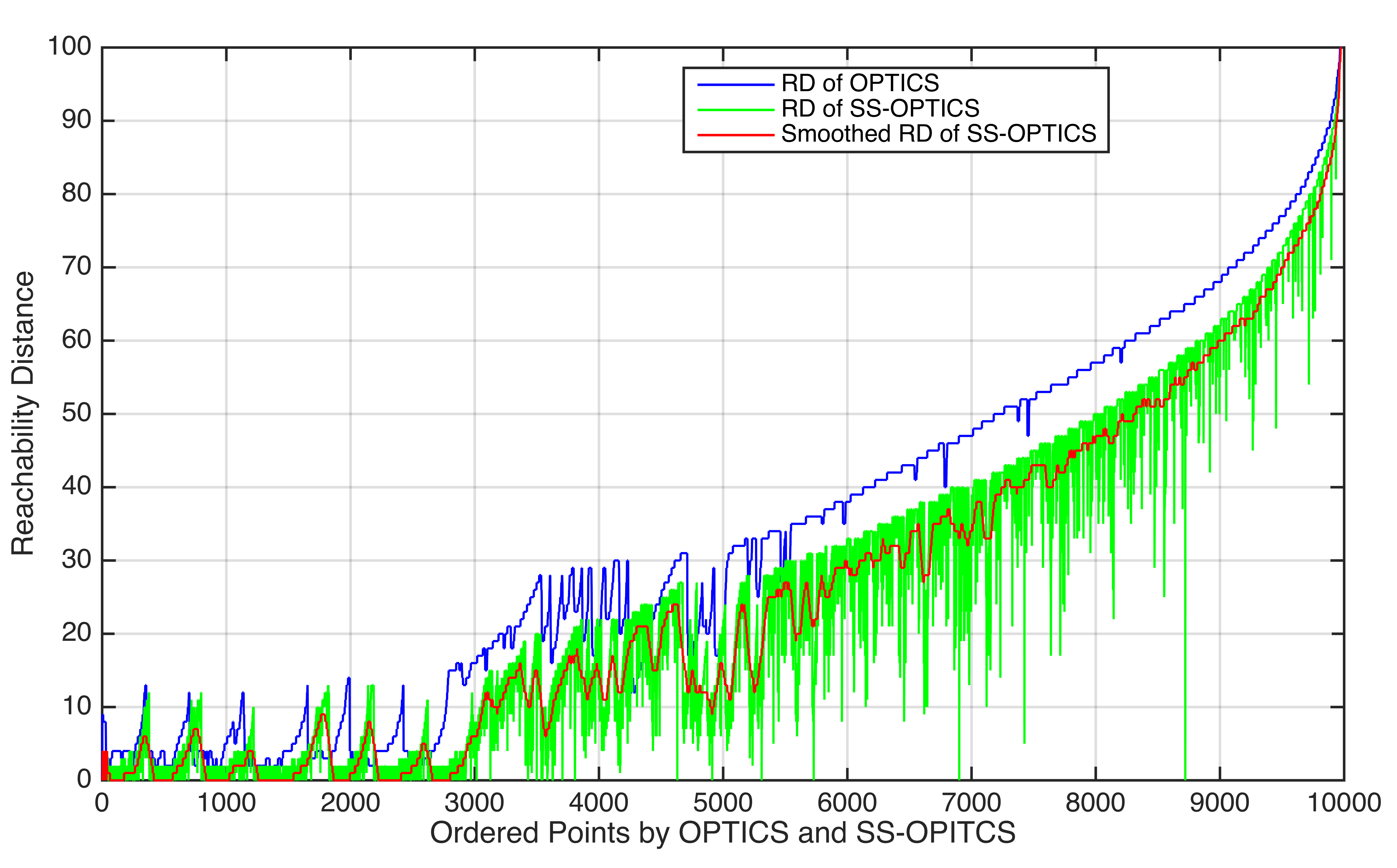}
\caption{RD curves of OPTICS and SS-OPTICS, $\varepsilon$ = 100 and S = 41}
\label{figure:RDcurve}
\end{figure}

\emph{1) Simplify}: The original OPTICS algorithm has two key concepts, \emph{cord-distance} and \emph{reachability-distance}. 

\textbf{Definition-1, \emph{$\varepsilon$-Neighborhood}}: Let $p$ be an object from a dataset D. The $\varepsilon$-neighborhood set of a point $p$ is defined by $N_{\varepsilon} (p) =\{x \in D | dist(p, x) \leq \varepsilon \}$.

\textbf{Definition-2, \emph{Core-distance (cd)}}:
Let $p$ $\in$ D, let $\varepsilon$ be a distance value, let $N_{\varepsilon}(p)$ be the $\varepsilon$ -neighborhood of $p$, let \emph{MinPts} be a natural number and let \emph{MinPts-distance(p)} be the distance from $p$ to its \emph{MinPts} neighbor. Then, the \emph{core-distance} of $p$ is defined as \emph{core-$distance_{\varepsilon,MinPts}$(p)} =
$$ \left\{
\begin{aligned}
&\emph{UNDEFINED}  \qquad ,\ if\ Card(N_{\varepsilon}(p)) < MinPts \\
&\emph{MinPts-distance} (\emph{p})  \qquad,\ otherwise
\end{aligned}
\right.
$$

\textbf{Definition-3, \emph{Reachability-distance (rd)}}:
Let $p$, $o$ $\in$ D, let $N_{\varepsilon}(o)$ be the $\varepsilon$ -neighborhood of $o$, let \emph{MinPts} be a natural number. Then, the \emph{reachability-distance} of $p$ with respect to $o$ is defined as \emph{reachability-$distance_{\varepsilon,MinPts}$(p,o)} =
$$ \left\{
\begin{aligned}
&\emph{UNDEFINED} \qquad ,\ if\ |N_{\varepsilon}(o)| < MinPts \\
&\emph{max}\ (\emph{core-distance}(\emph{o}),\ distance(\emph{o, p}))  \qquad,\ otherwise
\end{aligned}
\right.
$$
Here, $\varepsilon$ and \emph{MinPts} are two input parameters of the original OPTICS algorithm. 
According to OPTICS's definitions, the green points covered by the yellow circle in the Fig. \ref{figure:OPTICS} share the same reachability-distance (\emph{rd}), which equals to the core-distance of point \emph{o} (\emph{cd}). Although the green points, $p_{1}$, $p_{2}$, and $p_{3}$, have the same \emph{rd}, their actual reachable distances from point \emph{o} are different ($rd_{p_{1}}^{'} < rd_{p_{2}}^{'}<rd_{p_{3}}^{'}$). 

The main ideas of OPTICS can be described as 1) reachability distance represents density and 2) reachability-distance determines the points' output order, which determines clusters. Based on these ideas, we can find a design flaw of OPTICS that the output order of $p_{1}$, $p_{2}$, and $p_{3}$ in the left example of Fig. \ref{figure:OPTICS} maybe disordered due to their same \emph{rd}s. Thus, we design an improved OPTICS algorithm by abandoning the concept of \emph{core-distance} and define a new concept of \emph{reachability-distance (RD)} as follow.

\begin{figure*}[t]
\centering
\includegraphics[width=0.7\linewidth]{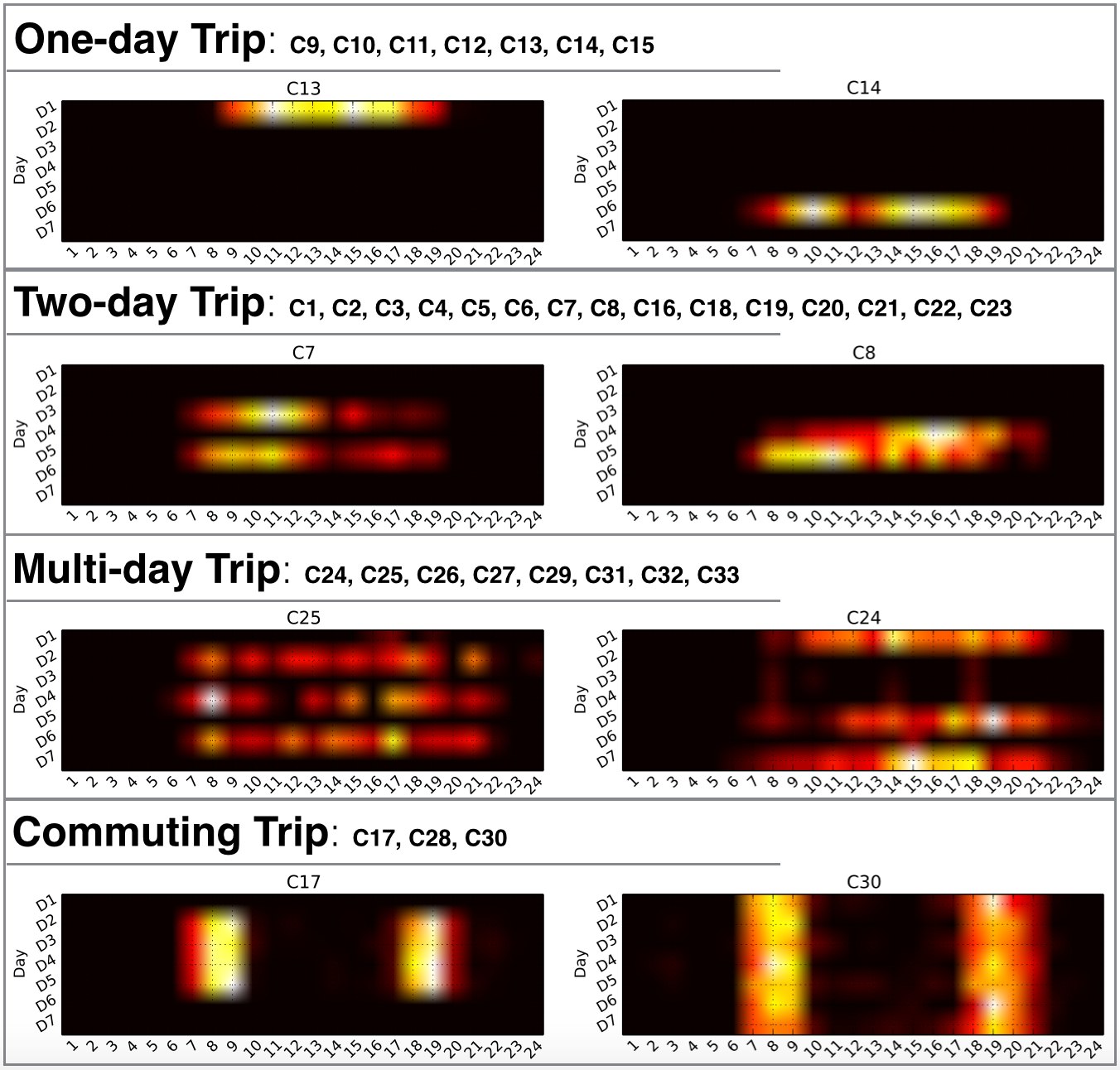}
\caption{4 obvious categories of the heatmap of the 33 clusters. D1-D5: weekdays, D6-D7: weekends}
\label{figure:heatmaps}
\end{figure*}

\IncMargin{1em}
\begin{algorithm}[!htbp]
 \KwData{D (Unprocessed Dataset), $\varepsilon$}
 \KwResult{OrderedPoints }
 initialization\;
 \While{$D \neq Null$}{
  $Point = D.pop()$\;
  $OrderedPoints.append(Point)$\;
  $P\_neighbors = point.neighbor(\varepsilon)$\;
  \If{$P\_neighbors \neq Null$}{
   $OrderSeeds$ = []\;
   $OrderSeeds.updateRD(Point, P\_neighbors)$\;
   \While{$OrderSeeds$}{
    $OrderSeeds.sort(key=RD)$\;
    $Seed = OrderSeeds.pop()$\;
    $OrderedPoints.append(Seed)$\;
    $S\_neighbors = Seed.neighbor(\varepsilon)$\;
    \If{$S\_neighbors \neq Null$}{
     $OrderSeeds.updateRD(Seed, S\_neighbors)$
    }}}}
 \caption{Getting Ordered Points by OPTICS}
\label{algorithm:OPTICS}
\end{algorithm}
\DecMargin{1em}

\textbf{Definition-4, \emph{New Reachability-distance (RD)}}:
Let $p$, $o$ $\in$ D, let $N_{\varepsilon}(o)$ be the $\varepsilon$ -neighborhood of $o$. The \emph{reachability-distance} of $p$ with respect to $o$ is defined as \emph{reachability-$distance_{\varepsilon}$(p,o)} =
$$ \left\{
\begin{aligned}
&\emph{UNDEFINED} \qquad\qquad\qquad\qquad ,\ if\ |N_{\varepsilon}(o)| = 0 \\
&\ distance(\emph{o, p}) \quad s.t.\quad p \in N_{\varepsilon}(o) \qquad,\ otherwise
\end{aligned}
\right.
$$
 
\emph{2) Smooth}: The 2D plot based on the ordered points' reachability distance can help us distinguish the clusters. As the denser the points gather, the lower reachability-distances the points get, the "valley" shapes in the reachability distance curve represent clusters with high density. In Fig. \ref{figure:RDcurve}, the blue line is the \emph{rd} curve of OPTICS, the green line is the \emph{RD} curve of SS-OPTICS. We notice that, although the average value of SS-OPTICS's RD is obviously less than OPTICS's, their curves are extremely similar.

The red line is the smoothed RD of SS-OPTICS, $RD^{'}$, in Fig. \ref{figure:RDcurve}.  We smooth the RD curve with two aims. One is to make it easier to identify the valley-shaped clusters, and the other is to control the size of a cluster. By using the mean filtering method to smooth the RD curve, we can achieve the two goals with only one parameter, window size (S). Each value of the smoothed RD curve, $RD_{i}^{'}$, is the mean of RD value of points within the window:

\begin{equation}
RD_{i}^{'} = (\sum_{j=i-n}^{j=i+n} RD_{j})/S, \quad s.t. \quad  n=\frac{S-1}{2}
\end{equation}
Since $RD^{'}$ has been filtered by a S sized window, it should be noticed that the boundary of the valley-shaped cluster has a bias to the left, and the offset is $\frac{S-1}{2}$. After the mean filtering, the valley (cluster) of the RD curve, whose number of the points in this cluster is less than $\frac{S-1}{2}$, will nearly be filled up. Thus, the cluster size is controlled to be larger than $\frac{S-1}{2}$.

Comparing to OPTICS, SS-OPTICS needs one parameter ($\varepsilon$) and OPTICS needs two ($\varepsilon$ and \emph{MinPts}). The time complexity of SS-OPTICS is $O(n^{2})$, same as OPTICS. Meanwhile, both the algorithms are not sensitive to the value of the parameters. The $\varepsilon$ is set to be 100. In addition, the SS-OPTICS is easier to control the cluster size and define the cluster boundary by setting the window size S. Since the window size S only affects the boundary of clusters, it does not influence the overall clustering performance. Thus, after experimental testing, the window size (S) is set to be 40 in this study. Finally, we iteratively cluster several random samples of SCD, containing 20000 entries in each sample, and identify 33 clusters for the next stage to classify the whole dataset. The sensitivity analysis on sample size shows that when the sample size is over 20000, the clustering results nearly converge, and thus, the sample size is set as 20000.

According to the transaction time distribution of the 33 clusters, they can be classified into 4 big categories obviously as Fig. \ref{figure:heatmaps} shows. The 4 categories can be described as: \emph{one-day trips}, \emph{two-days trips}, \emph{multi-days trips}, and \emph{commuting trips}. The one-day trips containing 7 clusters (9-15) are distributed in one day of the week from Monday to Sunday. The transaction time of two-day trips (cluster 1-8, 16 and 18-23) is distributed mainly in two days of the week, while the transaction time of multi-day trips (cluster 24-27, 29 and 31-33 ) is dispersed in different days (at least 3 days) of the week. The commuting trips (clusters 17, 28 and 30) are mainly charactered with regular morning and evening peaks during the week.

\begin{figure*}[t]
\centering
\includegraphics[width=0.8\linewidth]{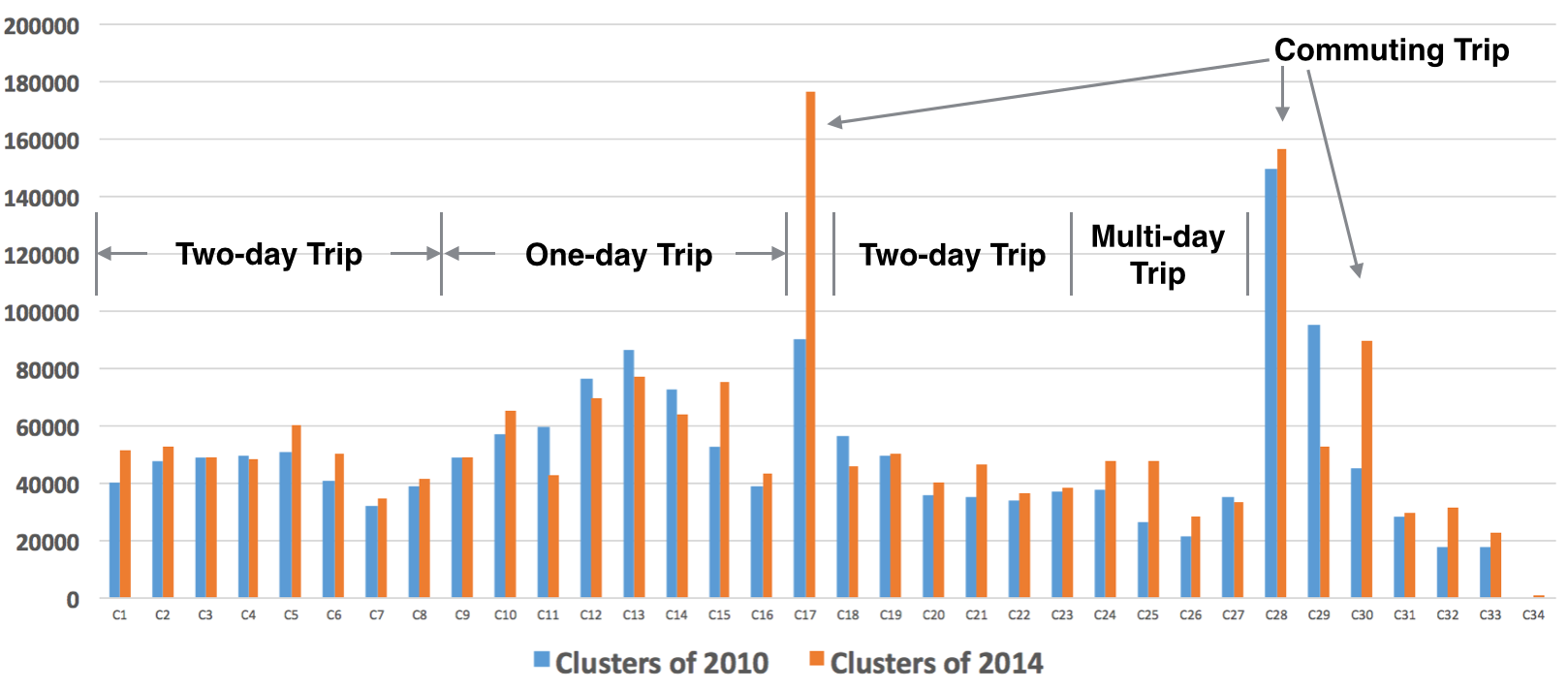}
\caption{Number of smart cards in each of the 34 clusters in 2010 and 2014}
\label{figure:clusterstatistics}
\end{figure*}

\subsection{Classifying SCD records based on SS-OPTICS results}

The set of the 33 clusters acquired by SS-OPTICS is denoted as $C=[C_{1},...,C_{33}]$. Each $C_{i}$ in $C$ is a vector containing 168 components, like the vector of smart card's transaction time. Each component ($c_{j}$) of cluster $C_{i}$ is the frequency of passengers' travel behavior in the ($j\%24$)th hour of the ($j-j\%24/7$)th day of the week. We also add a cluster to $C$ as the 34th cluster, whose components are all zero, to classify some noise points. Hence, taking the clusters' features as the centroids of clusters, $C=[c_{ij}]_{34\times168}$, all the SCD records can be grouped into 34 clusters.

According to the acquired data, two necessary aspects of information for clustering the whole dataset are obtained, i.e. 1) the cluster number $K=34$ and 2) the feature of each cluster $C_{i}$. With the two aspects of information, classifying the whole dataset can be easily fulfilled by utilizing the simple and efficient K-means algorithm. For each SCD vector, $\textbf{\emph{v}}=[v_0, \cdots , v_{167}]$, it belongs to Cluster $C_{i}$ satisfying
\begin{equation}
i = \arg\!\max_i \sum_{j=1}^{168}v_{j}\times c_{ij}
\end{equation}
Then, the cluster $C_{i}$ is updated such that $c_{ij}$=
$$ \left\{
\begin{aligned}
&\frac{n\times c_{ij}+v_{j}}{n+||\textbf{\emph{v}}||_{0}}\qquad ,\ if\ v_{j} \neq 0 \\
&\frac{n\times c_{ij}}{n+||\textbf{\emph{v}}||_{0}} \qquad \ \ \ ,\ otherwise
\end{aligned}
\right.
$$
Here, $n$ is the total number of transactions in $C_{i}$. After iteratively grouping all the SCD records, the clustering results are acquired and analyzed in the following section.

\subsection{Clustering Results and Analysis}

After the clusters were created for all data combined, the data sets of 2010 and 2014 can be classified accordingly. Then, we can get the numbers of the smart card records in each of the 34 clusters in the years of 2010 and 2014. Fig. \ref{figure:clusterstatistics} demonstrates the number of cards in each clusters. It can be noticed that the numbers of trips in \emph{one-day}, \emph{two-day} and \emph{multi-day} trips do not vary much over the four years from 2010 to 2014. The number of one-day trips, around 60000 in each cluster, is a little more than that of two-day trips, around 50000. Comparing to the clusters belonging to the multi-day trip group, the one-day and two-day trips occupy the most in the total trips. It reveals that more passengers in Beijing choose to use public transit occasionally, mainly in one day or two days. The number of multi-day trips (about 30000 in each cluster) is the least. This makes sense that fewer passengers would ride public transit vehicles or metros on multiple days in a week if they are not commuters. It should be noted that although special case, like misclassification, may exist in the clustering results, the overall analysis can hardly be influenced with help of the huge size of the data set. 

Almost all the towering bars in Fig. \ref{figure:clusterstatistics} belong to the commuting trip group. This group, whose travel patterns are shown in the commuting trip group in Fig. \ref{figure:heatmaps}, clearly represents the main members of public transit riders, namely the commuters who take a home-to-work trip in the morning and go back home in the evening every weekday. One interesting result is that the number of passengers belonging to commuting trip group nearly doubled from 2010 to 2014. There are two potential reasons leading to this huge increase of commuting trips in Beijing. One is that public transit became more convenience from 2010 to 2014. As we investigated, during this time period, Beijing metro constructed 8 more lines into 15 lines in total and the total metro length increased rapidly from 228 km to 465 km. The other reason is the ground transportation in Beijing became more congested and forced some people to choose public transit, since the total number of private vehicles in Beijing increased from 2.9 million in 2010 to 4.3 million in 2014.

\section{Mobility and Stability Analysis}

Mobility and stability patterns of people living in metropolitan areas are really hard to measure due to the huge number of residents and incomplete methods to probe all the population. As mentioned by many studies \cite{Ma20131,Mohamed2014Understanding,Pelletier2011557}, utilizing SCD collected by AFC system is a nearly ideal solution of this problem, since public transit is used by a large proportion of urban residents and AFC system can record their travel details. But we still need to consider the influence of many other factors, including residents age distribution, social scale, per capita income, type of job, city size and so on, to analyze transit passengers' travel behaviors. Since the datasets of 2010 and 2014 are selected according to same smart card IDs, the mobility and stability of fixed passengers can be reflected by the changes of their travel patterns between 2010 and 2014. Passengers' travel patterns are represented by the recorded transaction time when they using public transit. In this section, we analyze passenger's mobility and stability pattern based on temporal information combining some background socioeconomic factors listed in Table \ref{table:socialeconomics}.

\begin{figure}[!htbp]
\centering
\includegraphics[width=\linewidth]{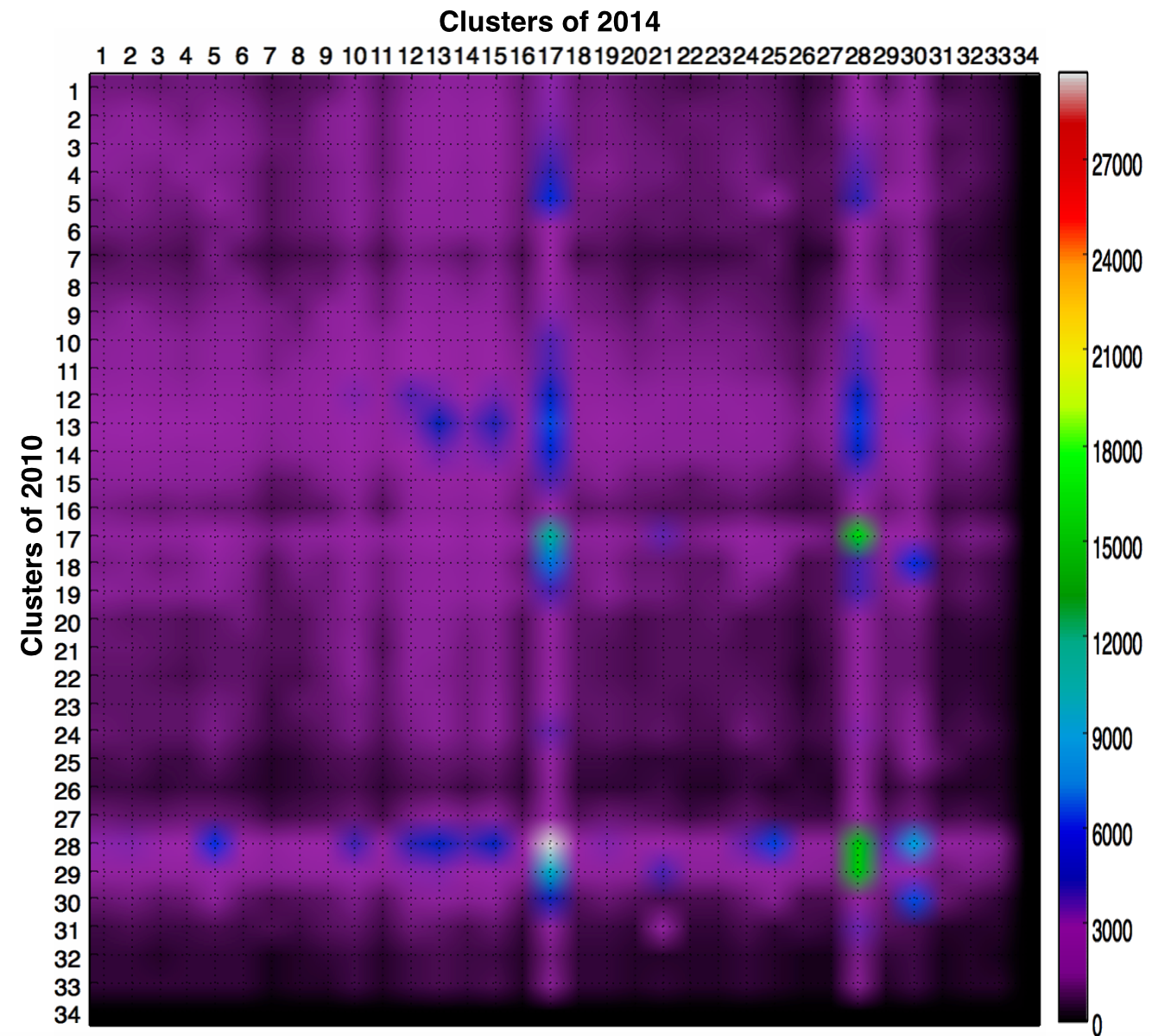}
\caption{Heatmap of the 34 clusters' transition matrix}
\label{figure:tansitionmatrix}
\end{figure}

\begin{figure*}[!htbp]
\centering
\includegraphics[width=0.8\linewidth]{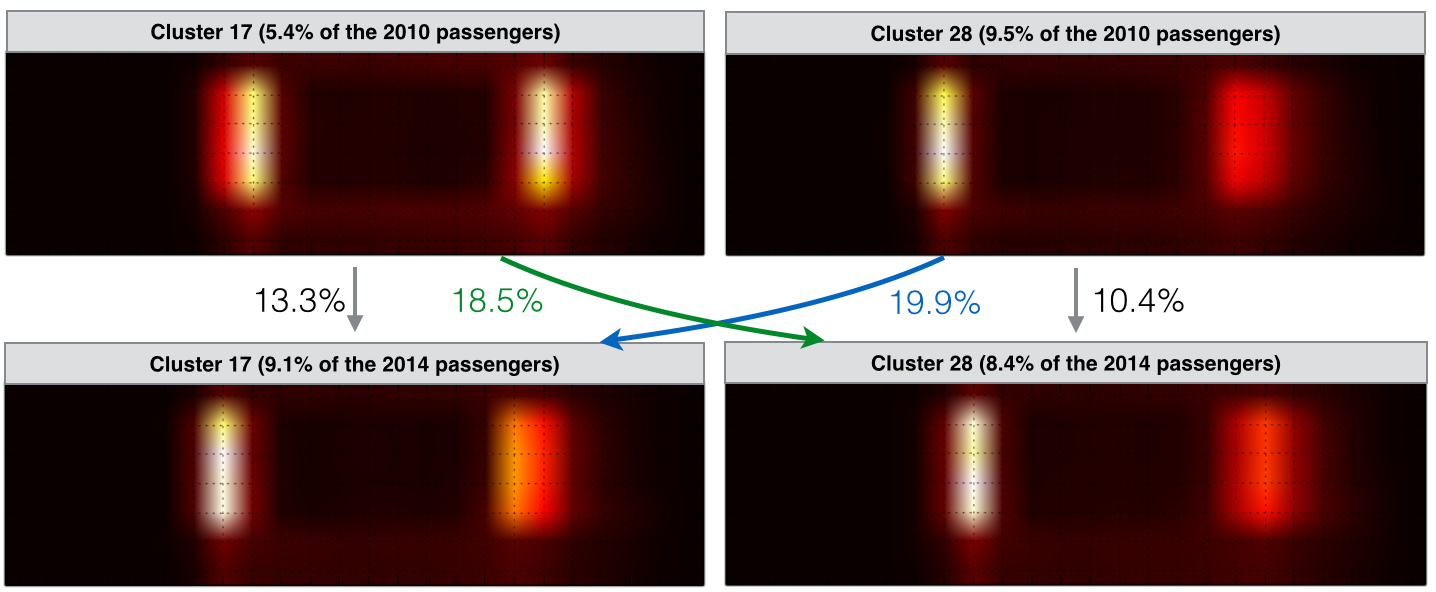}
\caption{Heatmaps of the mutual transition between cluster 17 and cluster 28 in both 2010 and 2014}
\label{figure:clustersample_17_28}
\end{figure*}

\begin{table*}[!htbp]
\begin{center}
\caption{Social Economics Factors in Beijing}
\label{table:socialeconomics}
%\rowcolors{2}{gray!25}{white}
\begin{tabular}{c|cccccccccc}
\toprule
%\rowcolor{gray!50}
$Year$ & $Population$ &  ${Population\ Density}$ & $Private\ Vehicles$ & $Bus\ Volume$ & $Metro\ Volume$ \\ 
        \midrule
 2010 & 17.55 $mil.$ & 1224  ${person/km^{2}}$ & 2.97 $mil.$ & 5.165 $bil.$ & 1.423 $bil.$  \\
 2014 & 21.15 $mil.$ & 1498  ${person/km^{2}}$ & 4.25 $mil.$ & 4.843 $bil.$ &  3.205 $bil.$ \\
\bottomrule
\end{tabular}
\end{center}
\end{table*}

\begin{table}[!htbp]
\begin{center}
\caption{Transition Matrix of Extreme Travelers}
\label{table:extremeTM}
\rowcolors{2}{gray!25}{white}
\begin{tabular}{cp{2em}p{2em}p{2em}p{1.5em}p{3.2em}|c}
\toprule
%\rowcolor{gray!50}
\diagbox[dir=SE,width = 4em]{2010}{2014} & EB & NO & TI & RI & NE & SUM\\ 
\midrule
EB & 1286 & 206 &	535 & 82 & 7605 & 9714\\ 
NO & 299 & 2550 &	2200 & 153 & 30006 & 35208\\ 
TI & 376 & 996 & 9488 & 182 & 48406 & 59448\\ 
RI & 93 & 198 & 677 & 275 &7351 & 8594\\
NE & 8780 & 26357 & 82630 & 3977 & 1646118 & 1767862\\ 
\midrule
SUM & 10834 & 30307 & 95530 & 4669 & 1739486 & 1880826\\ 
\bottomrule
\end{tabular}
\begin{tablenotes}
      \small
      \item EB: Early Birds, NO: Night Owls, TI: Tireless Itinerants, RI: Recurring Itinerants and NE: Non-Extreme Travelers
\end{tablenotes}
\end{center}
\end{table}

\subsection{Extreme Travelers Analysis}

According to the classification criteria proposed in Table \ref{table:extreme}, the transition matrix of the four types of extreme travelers (EB, NO, TI, RI) from 2010 to 2014 is generated and shown in Table \ref{table:extremeTM}. The numbers of extreme travelers in 2010 (141340) and 2014 (112964) are both very small compared to that of non-extreme travelers. In addition, 84\% of the extreme travelers in 2010 converted into non-extreme travelers in 2014, which means the stability of extreme travelers' live pattern cannot last for a long time. 

But among the four types of extreme travel patterns, it still can be found that the most passengers in the EB, NO, and TI groups in 2010, with the amounts of 1286, 2250 and 9488, staying in the same groups in 2014. Hence, that means passenger with an extreme travel pattern is more likely to keep the original travel pattern other than to convert into other extreme patterns. It also meets the findings of the previous work \cite{long2016early} that most of EB, NO, and TI are full-time workers, implying full-time worker will less likely change their jobs (also travel pattern) compared to the unemployed. 

\begin{table}[!h]
\begin{center}
\caption{Transition Matrix of Non-extreme Travelers}
\label{table:nonExtremeTM}
\rowcolors{2}{gray!25}{white}
\begin{tabular}{ccccc|c}
%\rowcolor{gray!50}
\toprule
\diagbox[dir=SE,width=4em]{2010}{2014} & O & T & M & C & SUM\\ 
\midrule
O & 119270 & 193290 &	84311 & 31864 & 428735\\
T & 164817 &	298667 &	142436 &	48043 & 653963\\
M & 64449 & 142399 & 76769 & 20038 & 303655\\
C & 36642 & 79266 & 47757 & 11812 & 175477\\
\midrule
SUM & 385178&	713622&	351273&	111757&	1561830\\
\bottomrule
\end{tabular}
\begin{tablenotes}
      \small
      \item O: One-day trip, T: Two-day trip, M: Multi-day Trip and C: Commuting Trip
\end{tablenotes}
\end{center}
\end{table}

\subsection{Non-Extreme Travelers Analysis}

\subsubsection{Fine-grained Analysis}
By acquiring the numbers of passengers of 34 clusters in 2010 and 2014, the transition (mobility) matrix of these clusters is calculated and demonstrated by a heatmap shown in Fig. \ref{figure:tansitionmatrix}. In this heatmap, the brighter the grid is, the more passengers belong to this grid. We can easily catch sight of brighter parts (green, blue and white part) and find them mainly distributed in cluster 17, cluster 28 and cluster 30 in both 2010 and 2014, which belong to the commuting trip category. 

Especially for the green and white grids ($C_{17\rightarrow17}$, $C_{17\rightarrow28}$, $C_{28\rightarrow17}$ and $C_{28\rightarrow28}$ ), the numbers of trips in these grids are several times larger than that of other grids. This reflects the travel pattern stability of the passengers belonging to the commuting trip category. These four grids' weekly profiles are demonstrated by heatmaps in Fig. \ref{figure:clustersample_17_28}. Although their morning and evening peak hours have a deviation of one hour, the stability can be reflected by the almost same trip occurrence time distribution and the same time intervals between morning and evening peak hours. Their temporal profiles also present most commuting trips of passengers in Beijing are distributed mainly from Tuesday to Friday. It is interesting to explore why commuting passengers tend to ride public transit on weekdays except for Monday. A possible explanation is the Monday Morning Syndrome (MMS), which means some people feel even more tired out on Monday than on Friday after the relaxation over the weekend. 

\begin{figure*}[t]
\centering
\includegraphics[width=0.8\linewidth]{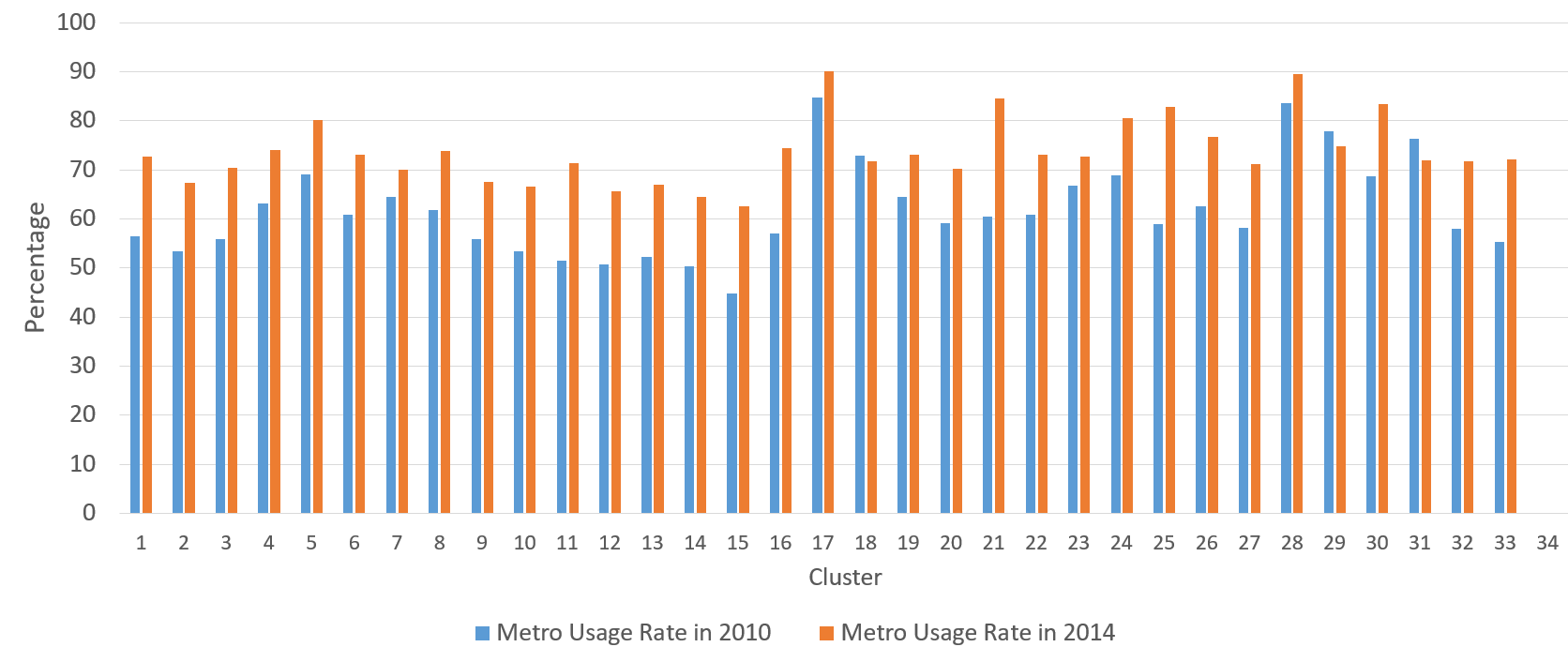}
\caption{Percentage of the passengers who take the metro at least once a week in each cluster}
\label{figure:bus_metroUsageRate}
\end{figure*}

There are also some blue grids distributed in the one-day trips region (cluster 9-15). The heatmap shows the mutual transitions between one-day trip category (cluster 9-15) and commuting trip category (cluster 17, 28) happen a lot. Passengers in the group of one-day trip category are regarded as the ones using public transit occasionally. This transition shows passengers change their public transit usage patterns from occasional to regular on weekdays. This situation can be the result of many reasons, like changing jobs or working locations, earning enough money to buy a car, or taking metro to work instead of driving. Fig. \ref{figure:bus_metroUsageRate} shows the percentage of passengers who rode a metro at least once a week in each cluster. The average percentage in 2014 is apparently higher than that of 2010. Further, as shown in the figure, the percentages of passengers in the commuting clusters (17, 28, and 30) reach the peaks in both lines of 2010 and 2014. This means commuting passengers may be the most stable group who are most willing to transit by the metro.

\subsubsection{Coarse-grained Analysis}
The transition matrix of the four groups of non-extreme travelers is also counted and shown in Table \ref{table:nonExtremeTM}. Each component of the transition matrix demonstrates the number of passengers transition from one group to another. Analyzing the transition between different groups provides a new perspective to analyze passenger's mobility and stability. However, only with smart card data, we cannot prove our conjectures. Hence, to better understand the mobility and stability of passengers, we combine socioeconomic statistics data of Beijing in both 2010 and 2014 \cite{Bureau2010Statistical}, \cite{Bureau2014Statistical}, shown in Table \ref{table:socialeconomics}. From 2010 to 2014, the population of Beijing increased by 3.6 million and the population density in the urban area rose from 1224 to 1498 persons per square kilometer. Along with the growth of population, the total number of private vehicles in Beijing increased from 2.97 million to 4.25 million. All these factors show that Beijing became more crowded in the urban area and more vehicles led more congested ground transportation after 2010. As for the transition matrix, the ratios of components in each row of the transition matrix are very close (approximately O:T:M:C=6:14:7:2), implying the overall travel patterns of passengers in Beijing did not change much from 2010 to 2014. Although the population and the number of vehicles increased a lot in Beijing, the travel patterns of public transit riders tend to be stable. However, as Table \ref{table:socialeconomics} indicates, the total volume of passengers riding metros doubled during the four years, while the volume of passengers taking buses decreased a little bit. This unusual decline might be the result of the rapid construction of the Beijing Metro System targeting at mitigating congestion brought by the increasing population and usage of private vehicles. However, as revealed by the transition matrix that the whole non-extreme travelers nearly keeps the same travel patterns. That means if the passengers' travel demand keeps at a similar level, constructing new metro lines may not be able to fundamentally solve the congestion problem.

\subsection{Discussion on Mobility and Stability}
Analyzing smart card data from different temporal scales can provide different points of view to understand the mobility and stability of transit passenger's travel behavior. The mobility and stability are relevant and A passenger's weekly travel records show his/her short-term mobility patterns, yet the change of whole passengers' mobility patterns over years may imply the unchangeable of their lifestyle or social status. Along with the increase of population, transit availability, and urban size in Beijing, inhabitant's travel pattern changes a lot, but the distributions of different types of trips nearly keep the same. This reveals that individuals' short-term mobility integrates together and forms the population's long-term stability. One interesting phenomenon is that the total mileage of Beijing metro doubled from 2010 to 2014, during which the number of commuting trips also nearly doubled, as shown in Fig. \ref{figure:clusterstatistics}. This cannot be the coincidence, since travel behaviors of inhabitants should be largely determined by the general environments, public infrastructure, and services of the city. The aforementioned fine-grained and coarse-grained comparisons of passengers' transit profiles between 2010 and 2014 both highlight trip category transitions between the 34 clusters and the four groups, which presents the distinctive long-term travel pattern dynamics.

\section{Conclusions}

Smart card data provide us new perspectives to observe the operation of our cities. In this paper, we analyze the temporal travel pattern of transit passengers in Beijing by clustering the SCD. To better analyze the SCD, we define a metric, i.e. transaction distance, to measure the similarity or difference between passengers' travel patterns by considering both time difference and frequency difference between SCD records. We also  propose a simplified smoothed OPTICS clustering method to cluster SCD. Comparing to the classical OPTICS methods, the SS-OPTICS needs fewer parameters and generates better clustering performance. We cluster the SCD based on different grouping granularities to analyze the long-term mobility and stability of transit passengers. By combining some socioeconomic data, we present several analyses about residents' temporal mobility and stability to elucidate the interdependence between mobility and stability of transit passengers' travel patterns. Extreme travelers are most vulnerable that the stability of extreme travelers' life pattern cannot last for a long time. According to clustering outcomes and our analyses, non-extreme travelers' high mobility is shown by the transition between different fine-grained clusters. However, the stability of their travel patterns is also obvious based on coarse-grained travel pattern categorization.

In the future study, the proposed transaction distance is very suitable for measuring the similarity of time series with specific physical meanings, like the passenger's travel record sequences in this study. Since the proposed SS-OPTICS algorithm can generate optimal clustering results, it can also be applied in transit analysis related applications. In addition, several improvements can be made based on the work presented herein. Firstly, the accuracy of SCD can be enhanced in the future by adopting robust methods to mitigate the deviations of boarding and alighting time. Secondly, the proposed SS-OPTICS algorithm can be improved aiming to find a better way to define the boundaries of clusters. Thirdly, more fine-grained socioeconomic and geospatial data can be incorporated in the analysis.

\section*{Acknowledgment}

This work is supported by National Natural Science Foundation of China (No.51408039).


\bibliographystyle{abbrv}
\bibliography{citation}

\begin{thebibliography}{10}

\bibitem{Ankerst1999OPTICS}
M.~Ankerst.
\newblock Optics : Ordering points to identify the clustering structure.
\newblock {\em Stanford Research Inst Memo Stanford University}, 28(2):49--60,
  1999.

\bibitem{Bagrow2012Mesoscopic}
J.~P. Bagrow and Y.-R. Lin.
\newblock Mesoscopic structure and social aspects of human mobility.
\newblock {\em Plos One}, 7(5):e37676, 2012.

\bibitem{briand2017analyzing}
A.-S. Briand, E.~C{\^o}me, M.~Tr{\'e}panier, and L.~Oukhellou.
\newblock Analyzing year-to-year changes in public transport passenger
  behaviour using smart card data.
\newblock {\em Transportation Research Part C: Emerging Technologies},
  79:274--289, 2017.

\bibitem{Bureau2010Statistical}
C.~S.~S. Bureau.
\newblock Statistical yearbook of china 2010.
\newblock {\em Bureau, China. State Statistical}, 2010.

\bibitem{Bureau2014Statistical}
C.~S.~S. Bureau.
\newblock Statistical yearbook of china 2014.
\newblock {\em Bureau, China. State Statistical}, 2014.

\bibitem{DBSCAN1996Density}
DBSCAN.
\newblock Density-based spatial clustering of applications with noise.
\newblock {\em Proceedings of International Conference on Knowledge Discovery
  and Data Mining}, 1996.

\bibitem{Hanson2005Perspectives}
S.~Hanson.
\newblock Perspectives on the geographic stability and mobility of people in
  cities.
\newblock {\em Proceedings of the National Academy of Sciences},
  102(43):p{\'a}gs. 15301--15306, 2005.

\bibitem{he2018classification}
L.~He, B.~Agard, and M.~Tr{\'e}panier.
\newblock A classification of public transit users with smart card data based
  on time series distance metrics and a hierarchical clustering method.
\newblock {\em Transportmetrica A: Transport Science}, pages 1--20, 2018.

\bibitem{Mohamed2014Understanding}
M.~E.~M. (Ifsttar, E.~C. (Ifsttar, J.~B. (Ifsttar, L.~O. (Ifsttar, and E.~C.
  (Ifsttar.
\newblock Understanding passenger patterns in public transit through smart card
  and socioeconomic data.
\newblock {\em Acm Sigkdd Workshop on Urban Computing}, 2014.

\bibitem{Kang2013Exploring}
C.~Kang, S.~Sobolevsky, Y.~Liu, and C.~Ratti.
\newblock Exploring human movements in singapore: a comparative analysis based
  on mobile phone and taxicab usages.
\newblock {\em Exploring human movements in Singapore: a comparative analysis
  based on mobile phone and taxicab usages - ResearchGate}, 2013.

\bibitem{kieu2015modified}
L.-M. Kieu, A.~Bhaskar, and E.~Chung.
\newblock A modified density-based scanning algorithm with noise for spatial
  travel pattern analysis from smart card afc data.
\newblock {\em Transportation Research Part C: Emerging Technologies},
  58:193--207, 2015.

\bibitem{Lathia2013Individuals}
N.~Lathia, C.~Smith, J.~Froehlich, and L.~Capra.
\newblock Individuals among commuters: Building personalised transport
  information services from fare collection systems.
\newblock {\em Pervasive and Mobile Computing}, 9(5):643--664, 2013.

\bibitem{long2016early}
Y.~Long, X.~Liu, J.~Zhou, and Y.~Chai.
\newblock Early birds, night owls, and tireless/recurring itinerants: An
  exploratory analysis of extreme transit behaviors in beijing, china.
\newblock {\em Habitat International}, 57:223--232, 2016.

\bibitem{long2015profiling}
Y.~Long and Z.~Shen.
\newblock Profiling underprivileged residents with mid-term public transit
  smartcard data of beijing.
\newblock In {\em Geospatial analysis to support urban planning in Beijing},
  pages 169--192. Springer, 2015.

\bibitem{Long2015Urban}
Y.~Long, L.~Sun, and T.~Sui.
\newblock A urban research literature review based on public transit smart card
  data (in chinese).
\newblock {\em Urban Planning Forum}, 3, 2015.

\bibitem{ma2017understanding}
X.~Ma, C.~Liu, H.~Wen, Y.~Wang, and Y.-J. Wu.
\newblock Understanding commuting patterns using transit smart card data.
\newblock {\em Journal of Transport Geography}, 58:135--145, 2017.

\bibitem{Ma20131}
X.~Ma, Y.-J. Wu, Y.~Wang, F.~Chen, and J.~Liu.
\newblock Mining smart card data for transit riders' travel patterns.
\newblock {\em Transportation Research Part C: Emerging Technologies}, 36(0):1
  -- 12, 2013.

\bibitem{mohamed2017clustering}
K.~Mohamed, E.~C{\^o}me, L.~Oukhellou, and M.~Verleysen.
\newblock Clustering smart card data for urban mobility analysis.
\newblock {\em IEEE Transactions on Intelligent Transportation Systems},
  18(3):712--728, 2017.

\bibitem{Noulas2012A}
A.~Noulas, S.~Scellato, R.~Lambiotte, M.~Pontil, and C.~Mascolo.
\newblock A tale of many cities: Universal patterns in human urban mobility.
\newblock {\em Plos One}, 7(5):: e37027., 2012.

\bibitem{Pelletier2011557}
M.-P. Pelletier, M.~Tr{\'e}panier, and C.~Morency.
\newblock Smart card data use in public transit: A literature review.
\newblock {\em Transportation Research Part C: Emerging Technologies},
  19(4):557 -- 568, 2011.

\bibitem{Peng2012Collective}
C.~Peng, X.~Jin, K.-C. Wong, M.~Shi, and P.~Li{\`o}.
\newblock Collective human mobility pattern from taxi trips in urban area.
\newblock {\em Plos One}, 7(4):: e34487., 2012.

\bibitem{Rodriguez2014Clustering}
A.~Rodriguez and A.~Laio.
\newblock Clustering by fast search and find of density peaks.
\newblock {\em Science}, 344(6191):1492--1496, 2014.

\bibitem{Uppoor2014Generation}
S.~Uppoor, O.~Trullols-Cruces, M.~Fiore, and J.~M. Barcelo-Ordinas.
\newblock Generation and analysis of a large-scale urban vehicular mobility
  dataset.
\newblock {\em IEEE Transactions on Mobile Computing}, 13(5):1--1, 2014.

\bibitem{Veloso2011Sensing}
M.~Veloso, S.~Phithakkitnukoon, and C.~Bento.
\newblock Sensing urban mobility with taxi flow.
\newblock In {\em Proceedings of the 3rd ACM SIGSPATIAL International Workshop
  on Location-Based Social Networks}, 2011.

\bibitem{zhang2016spatiotemporal}
F.~Zhang, J.~Zhao, C.~Tian, C.~Xu, X.~Liu, and L.~Rao.
\newblock Spatiotemporal segmentation of metro trips using smart card data.
\newblock {\em IEEE Transactions on Vehicular Technology}, 65(3):1137--1149,
  2016.

\bibitem{zheng2014efficient}
B.~Zheng, K.~Zheng, M.~A. Sharaf, X.~Zhou, and S.~Sadiq.
\newblock Efficient retrieval of top-k most similar users from travel smart
  card data.
\newblock In {\em 2014 IEEE 15th International Conference on Mobile Data
  Management}, volume~1, pages 259--268. IEEE, 2014.

\bibitem{Zheng2014Urban}
Y.~Zheng, L.~Capra, O.~Wolfson, and H.~Yang.
\newblock Urban computing: Concepts, methodologies, and applications.
\newblock {\em Acm Transactions on Intelligent Systems and Technology},
  5(3):222--235, 2014.

\end{thebibliography}
%
% once the .bbl file has been generated then place the text in your article.

% To get the numbered reference style the author should use [numbib]
%as an option in the document class.  For example: \documentclass[numbib]{imaiai}
%
%\begin{thebibliography}{9}
%\bibitem{1}
%Vicsek T., Czir\'{o}k A., Ben-Jacob E., Cohen I.: `Novel type of phase transition in a system of self-driven particles', \textit{Phys. Rev. Lett.}, 1995, \textbf{75}, pp.~1226--1229
%
%\bibitem{2}
%Jadbabaie A., Lin J., Morse A.S.: `Coordination of groups of
%mobile autonomous agents using nearest neighbor rules', \textit{IEEE
%Trans. Autom. Control}, 2003, \textbf{48}, pp.~988--1001
%
%\bibitem{3}
%Olfati-Saber R., Murray R.M.: `Consensus problems in networks
%of agents with switching topology and time-delays', \textit{IEEE Trans.
%Autom. Control}, 2004, \textbf{49}, pp.~1520--1533
%\end{thebibliography}

\end{document}